\documentclass[aps,pre,twocolumn]{revtex4-1}
\usepackage{amsmath,amssymb,graphicx}
\input epsf
\usepackage{amsbsy}
\usepackage{latexsym}
\usepackage{color}
\usepackage{graphicx}
\usepackage{psfrag}
\usepackage[normalem]{ulem}

\begin{document}
\title{Transition to coarsening for confined one-dimensional interfaces with bending rigidity}
\author{Thomas Le Goff$^{1}$, Paolo Politi$^{2,3}$ and Olivier Pierre-Louis$^{1}$}
\email{olivier.pierre-louis@univ-lyon1.fr}
\affiliation
{$^{1}$Institut Lumi\`ere Mati\`ere, UMR5306 Universit\'e Lyon 1-CNRS, Universit\'e de Lyon 69622 Villeurbanne, France}
\affiliation
{$^{2}$Istituto dei Sistemi Complessi, Consiglio Nazionale delle Ricerche, 
Via Madonna del Piano 10, 50019 Sesto Fiorentino, Italy}
\affiliation{
$^{3}$INFN Sezione di Firenze, via G. Sansone 1, 50019 Sesto Fiorentino, Italy}
\date{\today}

\begin{abstract}
We discuss the nonlinear dynamics and fluctuations of interfaces with bending rigidity
under the competing attractions of two walls with arbitrary permeabilities.  
This system mimics the dynamics of confined membranes.
We use a two-dimension hydrodynamic model, where membranes are effectively
one-dimensional objects.
In a previous work~[T. Le Goff {\it et al}, Phys. Rev. E
90, 032114 (2014)], we have shown that this model predicts
frozen states caused by bending rigidity-induced oscillatory interactions between kinks (or domain walls).
We here demonstrate that in the presence of  tension, potential asymmetry,
or thermal noise, there is a finite threshold above which
frozen states disappear, and perpetual coarsening is restored.
Depending on the driving force, the transition to coarsening exhibits different scenarios.
First, for membranes under tension, small tensions can only lead to transient 
coarsening or partial disordering, while
above a finite threshold, membrane oscillations disappear
and perpetual coarsening is found.
Second, potential asymmetry is relevant in the non-conserved case only,
i.e. for permeable walls, where it induces a drift force on the
kinks, leading
to a fast coarsening process via kink-antikink annihilation. However,
below some threshold, the drift force can be balanced by the oscillatory
interactions between kinks, and frozen adhesion patches can still be observed.
Finally, at long times, noise restores coarsening with standard exponents
depending on the permeability of the walls.
However, the typical time for the appearance of coarsening exhibits an Arrhenius form.
As a consequence, a finite noise amplitude is needed in order 
to observe coarsening in observable time.
\end{abstract}

\maketitle



\section{Introduction}

Bending rigidity is a crucial ingredient in soft matter systems,
which leads to a number of nontrivial effects in the 
equilibrium and non-equilibrium behavior of membranes~\cite{Canham1970,Helfrich1973}
and filaments~\cite{Marko1995,Benetatos2003,PierreLouis2011}. For example, in equilibrium, the minimization of bending energy leads
to non-trivial shapes of membrane vesicles\cite{Lipowsky1991}, and knotted filaments\cite{Gallotti2007}.
In non-equilibrium conditions, bending rigidity also plays a crucial
role in cell adhesion, or in the rheology and stability of membrane stacks~\cite{Auth2013,Hemmerle2013}.

Here we explore the dynamical behavior of a model 
where  an interface with bending rigidity is confined between two walls.
This model aims at understanding the behavior 
of lipid membranes confined into double-well potentials. Such
double well potentials have been evidenced in various experimental contexts,
for example when a membrane is placed under the combined influences of a short-range attraction
induced by ligand-receptor pairs, and a medium range repulsion
caused by polymer brushes mimicking glycocalyx~\cite{Bruinsma2000,Sengupta2010}.
A second example is that of intermembrane attraction in the presence of ligand-receptor
pairs with two different lengths~\cite{Weikl2009}. We also expect
this double-well picture to represent membrane adhesion to the cytoskeleton
or to a substrate during cell adhesion.
Indeed, cell membranes are known to be able to bind to, or unbind from the cytoskeletal cortex
during the adhesion of cells to a substrate, e.g.
to form blebs~\cite{Brugues2010,Charras2008,Paluch2005}.

For the sake of simplicity, and following our previous work~\cite{LeGoff2014},
we consider a two-dimensional system.  As a consequence, 
the interface---hereafter denoted as a membrane---is effectively a one-dimensional object.
The walls attract the membrane, mimicking
physical adhesion~\cite{Swain2001,Hemmerle2013,Helfrich1978} (such as Van der Waals interactions,
hydration interactions, osmotic pressures, and entropic interactions), 
or specific adhesion~\cite{Zuckerman1995,Lipowsky1996}
via a simple effective adhesion potential. We have recently shown
that in such a model the membrane bending rigidity leads to 
arrested dynamics~\cite{LeGoff2014}, with 
frozen adhesion patches on both walls.

Other studies in the literature have suggested 
that finite-size adhesion patches may also be induced by more complex (bio)physical
ingredients, such as the clustering
ligand-receptor pairs~\cite{Lorz2007,ReisterGottfried2008,Bihr2012},
the disorder of the environment~\cite{Speck2012},
the trapping of ligands in membrane partitions~\cite{Kusumi2005},
or the active remodeling of the cytoskeleton~\cite{DelanoeAyari2004}.
One aim of our simple modeling 
is to provide hints towards a better understanding of the
formation of finite-size adhesion patches.

Moreover, the spatial organization of the frozen states 
observed in Ref.~\cite{LeGoff2014}
was strongly influenced by the permeability of the confining walls.
Indeed, strong wall permeabilities were shown to give rise to 
disordered states, while vanishing permeabilities lead to ordered states
with a periodic arrangement of patches~\cite{LeGoff2014}.

Here, we discuss the robustness of the frozen states
with respect to various physical  ``perturbations":
membrane tension, potential asymmetry,
and thermal fluctuations.
We find that in all cases, the frozen states can be
destroyed, and coarsening can be restored when the amplitude
of these effects exceeds a finite threshold.

In the following, we start in Section~\ref{s:model} with the derivation of the 
evolution equation for the membrane in the lubrication limit.
We focus on the two opposite regimes of very permeable, and perfectly
impermeable walls, respectively leading to non-conserved and conserved dynamics.
We then report the equations of  quantitative kink model derived in Ref.~\cite{LeGoff2015a}.
We also recall the main results regarding the existence 
of frozen states~\cite{LeGoff2014} in Section~\ref{s:recall}.

In Section~\ref{s:tension}, we show that there is a critical
tension $\sigma_c$ above which the oscillations of the membrane profile
disappear. In this regime, the resulting dynamics is similar to
that of the standard Time-Dependent Ginzburg-Landau (TDGL) equation~\cite{Hohenberg1977}
for the permeable case, and Cahn-Hilliard (CH) equation~\cite{Hohenberg1977,Cahn1958} for the
impermeable case, with perpetual coarsening caused by attractive
interactions between neighboring kinks.
For finite tensions below the threshold $\sigma_c$,
one can observe transient coarsening (i.e. which stops
after some finite time). This transient coarsening
is able to alter the perfect order observed
in the conserved case in the absence of tension. 

Then, in Section~\ref{s:asymmetry}, we discuss the consequences
of an asymmetric adhesion potential, favoring the
adhesion on one of the two walls.
While it is irrelevant in the conserved case,
this asymmetry gives rise to a
drift force on the kinks in the non-conserved case. 
This drift tends to increase
the size of the adhesion patches on the favored wall.
When the asymmetry is large enough, the resulting drift
leads an enhanced kink-antikink collision and annihilation rate, 
giving rise to a fast coarsening scenario with a final
state where the membrane is only on the side of the favored wall.
However, for weak asymmetries, the drift force on the kinks
is not strong enough to overcome kink interactions.
Thus, localized frozen adhesion patches 
can still be found.

Furthermore, the effect of thermal noise is analyzed in Section~\ref{s:noise},
using the kink model supplemented with Langevin forces. The results indicate
that noise always lead to perpetual coarsening. However,
the time for the appearance of coarsening exhibits
and Arrhenius law. Hence, observable coarsening
in a finite time requires a finite noise amplitude,
i.e. a finite temperature.

Finally, we conclude in Section~\ref{s:conclusion}.


\section{Membrane lubrication model}
\label{s:model}

\begin{figure}
\includegraphics[width=1.\columnwidth]{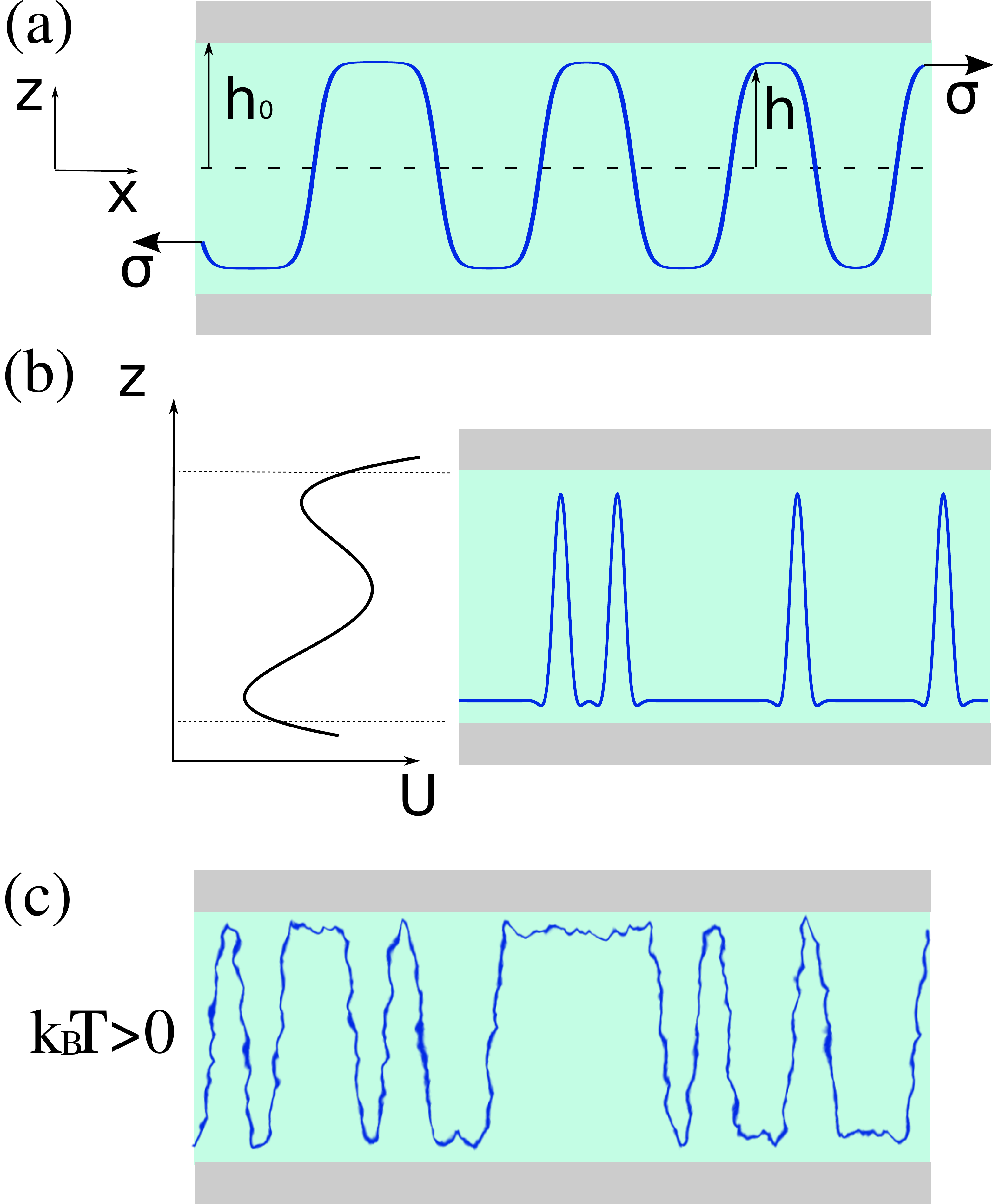}
\caption{Schematic of a confined membrane. 
The wall on the top is at $z=h_{0}$,
and the bottom wall is at $z=-h_{0}$. 
The blue solid line at $h(x,t)$ is the height of the membrane. 
Quantities above the membrane are written with $+$ and those below with $-$.
We discuss the influence of three physical ingredients
on the dynamics of confined membranes (a) a tension $\sigma$,
(b) an asymmetry of the adhesion potential ${\cal U}(h)$,
and (c) thermal fluctuations.}
\label{schemasysteme}
\end{figure}

Assuming for simplicity a two-dimensional system
in the $z,x$ plane, we wish to describe the dynamical behavior of a lipid membrane of height $z=h(x,t)$.
This membrane is confined between two walls at $z=\pm h_{0}$, as shown in Figure~(\ref{schemasysteme}),
and is surrounded by an incompressible fluid in the Stokes regime, obeying 
\begin{eqnarray}
{\nabla}p-\mu\nabla^2\mathbf{v}=0.
\end{eqnarray}
Here, $p$ is the pressure, $\mu$ the viscosity, and $\mathbf{v}$ 
is the fluid velocity. The membrane is assumed to be
impermeable. Moreover we consider a no slip condition 
at the walls, leading to $v_x(\pm h_0)=0$. Arbitrary wall permeability 
is accounted for using a phenomenological kinetic law
\begin{eqnarray}
v_{z\pm}(\pm h_{0})=\pm\nu(p_{\pm}-p_{ext}),
\end{eqnarray}
where $\nu$ is permeability kinetic coefficient,
$p_\pm$ is the pressure at $z=\pm h_0$,
and $p_{ext}$ is a constant external pressure.

Here, we consider both membrane bending rigidity and membrane tension 
via the standard Helfrich model~\cite{Canham1970,Helfrich1973}.
The membrane is also subject to a double well potential ${\cal U}(h)$,
which accounts for its interaction with the walls.
In the small slope limit, which is 
discussed below, the total energy of the membrane is
\begin{eqnarray}
\mathcal{E}=\int\mathrm{d}x~\left(\frac{\kappa}{2}(\partial_{xx}h)^{2}+\frac{\sigma}{2}(\partial_{x}h)^{2}+\mathcal{U}(h)\right).
\end{eqnarray}
where $\partial_x$ denotes the partial derivative with 
respect to $x$,  
$\kappa$ is the bending rigidity modulus, and $\sigma$ is the tension.

Following Ref.~\cite{LeGoff2014}, we consider the 
lubrication regime, where
the horizontal scale is much bigger than vertical scale, i.e.  $\partial_{x}h\ll1$
with $h\sim h_0$. In this limit, the hydrodynamic flow is to leading order
along $x$, with a parabolic Poiseuille-like profile along $z$.
In addition, we focus on the two
limits of very permeable walls $\nu\rightarrow\infty$, and
impermeable walls $\nu\rightarrow 0$, respectively leading to
\begin{eqnarray}
\partial_{t}h&=&\frac{\nu}{2}f_{z},
\label{e:permeable}
\\
\partial_{t}h&=&\partial_{x}\left[
-\frac{h_{0}^{3}}{24\mu}
\left(1-\frac{h^{2}}{h_{0}^{2}}\right)^{3}
\partial_{x}f_z\right],
\label{e:impermeable}
\end{eqnarray}
with the force along $z$ acting on the membrane
\begin{eqnarray}
f_{z}=-\kappa\partial_{x}^{4}h+\sigma\partial_{xx}h-{\cal U}'(h).
\label{e:force}
\end{eqnarray}
As discussed in Ref.~\cite{LeGoff2014}, in the impermeable case
there are additional nonlocal terms. However, these terms are irrelevant for
dynamics. We can therefore safely neglect them.
For the sake of simplicity
we also neglect the nonlinear mobility term $(1-h^2/h_0^2)^3$ appearing
in the conserved case. Its main consequence is
to slow down the dynamics in the late stages by a constant factor. Finally,
in rescaled coordinates, Eqs.(\ref{e:permeable},\ref{e:impermeable}) now read
\begin{eqnarray}
\label{e:TDGL4tension}
\partial_{T}H&=&-\partial_{X}^{4}H+\gamma\partial_{XX}H-U'(H),
\\
\label{e:CH4tension}
\partial_{T}H&=&\partial_{XX}\left(\partial_{X}^{4}H-\gamma\partial_{XX}H+U'(H)\right),
\end{eqnarray}
with  $X=[\mathcal{U}_{0}/(\kappa h_0^2)]^{1/4}x$ and $H=h/h_0$.
The normalized time variable is 
$T=t\nu\mathcal{U}_{0}/(2h_0^2)$ in the non-conserved permeable case,
and $T=\mathcal{U}_{0}^{3/2}t/(24\mu\kappa^{1/2})$ in the conserved impermeable case.
We have also defined the normalized tension:
\begin{eqnarray}
\gamma=\frac{h_{0}\sigma}{\kappa^{{1}/{2}}\mathcal{U}_{0}^{{1}/{2}}}.
\end{eqnarray}
Using the parameter $\gamma$, two limits can be defined.
First, when $\gamma\gg 1$, i.e. $\kappa\rightarrow 0$, the fourth order
derivative in the force term in Eqs.~(\ref{e:TDGL4tension},\ref{e:CH4tension}) is negligible,
and one recovers the standard Time Dependent Ginzburg Landau (TDGL) equation,
and Cahn Hilliard (CH) equation respectively. In the opposite
case $\gamma\ll 1$, i.e. $\sigma\rightarrow 0$, the second order derivative in the force term
in Eqs.~(\ref{e:TDGL4tension},\ref{e:CH4tension}) is negligible.
We denote the resulting equations as the TDGL4 and CH4 equations respectively.

An additional level of coarse graining is possible,
based on the dynamics of kinks, which are defined
as the transition zones, or domain walls, separating adhesion patches
in the two different wells of the potential $U$.
Kink dynamics have been derived in the 80s for the TDGL and CH equations~\cite{Kawasaki1982,Kawakatsu1985}.
Following our recent generalized derivation of kink dynamics
including bending rigidity~\cite{LeGoff2015a}, and a symmetric potential
($U(-H)=U(H)$) the position $X_n$ of the $n$th kink obeys:
\begin{eqnarray}
\dot{X}_n&=&\frac{1}{B_1}\Delta \tilde{R}_{n},
\label{e:kink_dynamics_TDGL_ext}
\\
\dot{X}_n&=&\frac{1}
{B_0^2\ell_{n-1/2}\ell_{n+1/2}-B_2(\ell_{n-1/2}+\ell_{n+1/2})} \Big\{
\nonumber \\
&&
\ell_{n-1/2}\left(B_2\dot{X}_{n+1}+\tilde{R}_{n+3/2}-\tilde{R}_{n-1/2}\right)
\nonumber \\
&&
+\ell_{n+1/2}\left(B_2\dot{X}_{n-1}+\tilde{R}_{n+1/2}-\tilde{R}_{n-3/2}\right)
\Big\}
\nonumber \\
\label{e:kink_dynamics_CH_ext}
\end{eqnarray}
for permeable and impermeable walls, respectively.
Here above the most important quantity is
the function $\tilde{R}(\ell)$, derived in Ref.~\cite{LeGoff2015a}, which is exponentially decreasing
and possibly oscillating for large $\ell$. Its explicit form will be given later.
As for notations, we have defined the difference
operator $\Delta$ such that $\Delta Y_n=Y_{n+1/2}-Y_{n-1/2}$
for any quantity $Y_n$. Moreover, the inter-kink distance
is denoted as $\ell_{n+1/2}=X_{n+1}-X_n=\Delta X_{n+1/2}$, 
and we also use the notation $\tilde{R}_{n+1/2}=\tilde{R}(\ell_{n+1/2})$. 
Finally,  the constants
\begin{eqnarray}
B_0&=& H_k(+\infty)-H_k(-\infty),
\\
B_1&=&\int_{-\infty}^{+\infty}\!\!\!\!\mathrm{d}X~(\partial_{X}H_k)^{2},
\\
B_2&=&\int_{-\infty}^{\infty}\!\!\!\!\mathrm{d}X~[H_k(+\infty)-H_k(X)][H_k(X)-H_k(-\infty)],
\nonumber \\
\end{eqnarray}
are calculated from the profile $H_k(X)$ of an isolated kink.
Notice that $B_0$ is simply the distance between the two minima
of the double well potential. $B_0>0$ for kinks,
and $B_0<0$ for antikinks.  In addition, we have $B_1>0$.
Moreover, $B_2>0$ is positive for all monotonically
varying kink profiles. For more complex profiles, 
the sign of $B_2$ is not know a priori, but
for all cases discussed below $B_2>0$.

The implementation of the kink dynamics provides an analytically
simplified---and numerically lighter---way to compute membrane dynamics.
The quantitative accuracy of these equations was tested and confirmed in Ref.~\cite{LeGoff2015a}
from a direct integration of the full dynamics.
Kink dynamics is asymptotically exact when the distance between kinks is large.


\section{The basic model: Frozen states and order-disorder transition}
\label{s:recall}

In Ref.~\cite{LeGoff2014} we have studied the dynamics
emerging from Eqs.(\ref{e:TDGL4tension},\ref{e:CH4tension}) in the absence of tension $\sigma=0$,
for a symmetric potential ${\cal U}(h)={\cal U}(-h)$,
and without fluctuations. Let us recall the main results.

First, both in the permeable and impermeable cases,
dynamics are rapidly arrested and 
the membrane profile reaches a frozen steady-state.
The origin of this steady-state was traced back to the 
presence of oscillatory interactions between kinks,
which are caused by the bending rigidity.

Using typical orders of magnitude with a physical 
potential including hydration repulsion and 
van der Waals attraction~\cite{Swain2001},
we find that the  scale $L$ of the adhesion
patches is~\cite{LeGoff2014} $L\sim h_0^{1/2}\kappa^{1/4}{\cal U}_0^{-1/4}$. 
Using $h_{0}\approx10$ to $20$nm as suggested by experiments in Refs.~\cite{Bruinsma2000,Weikl2009}, 
typical adhesion patch lengths are predicted to range from $100$~nm to $1~\mu$m.
Since the slopes are bounded at all times in the rescaled
coordinates, they remain small in physical coordinates, and the 
lubrication approximation is self-consistent.

The second main result of Ref.~\cite{LeGoff2014}
is that, starting from small random initial conditions, 
very permeable or impermeable walls respectively lead to disordered or ordered 
frozen configurations. 
In order to understand this result,
we first recall that the membrane is initially destabilized 
by the competitive attractions of the two walls in opposite directions.
Since short wavelength perturbations
are stabilized by bending rigidity, the instability
can only be present at long enough wavelengths.
For permeable walls, modes of large wavelength
all have the same dissipation rate, which is essentially
that of the translation of a flat membrane in the $z$ direction.
As a consequence, all the long wavelength modes have the same growth rate,
and many wavelengths are present simultaneously,
leading to a disordered membrane profile.
However, for impermeable walls, this translational mode along $z$
is forbidden due to mass conservation, and the increase of the
amplitude of long wavelengths modes
require flows along $x$ over large scales which are impeded 
due to their large cost in viscous dissipation. 
Since long wavelength perturbations are slowed down,
and small wavelength perturbations are stable, 
an optimum wavelength exists and
the instability develops at some well-defined intermediate
scale, leading to an ordered periodic state.

This result is readily obtained from the linear stability
analysis of Eqs.(\ref{e:TDGL4tension},\ref{e:CH4tension}). Indeed,
assuming small perturbations of amplitude $\epsilon\ll 1$, and  wavelength $\lambda$ 
around a flat profile $H=\bar{H}+\epsilon\exp i(\omega T+qX)$,
with $q={2\pi}/{\lambda}$,
one obtains the following dispersion relation:
\begin{eqnarray}
i\omega=- A_q [q^{4}+\gamma q^{2}+U''(\bar{H})],
\end{eqnarray} 
where  we have defined $A_q=1$ for the non-conserved case and $A_q=q^2$ for the conserved case. 
A positive real part of the growth rate $i\omega$ indicates an instability. Thus,
the membrane is unstable when  $U''(\bar{H})\leq0$. 
In addition, we confirm that all long wavelength perturbations grow with the same growth rate
in the case of TDGL4.
In contrast, the growth rate $i\omega$ exhibits
a maximum for CH4.

From the analysis of the periodic nonlinear steady-states, 
we have further shown in Ref.~\cite{LeGoff2014} 
that the periodic pattern emerging  in the impermeable 
case are stable, so that no further evolution is possible and 
the membrane profile remains frozen in this ordered state.
In the non-conserved case, the disordered
pattern emerging from the linear instability rearranges,
and reaches a frozen disordered steady-state~\cite{LeGoff2014}.

We have now finished to review the basic model.
The next three Sections report original results on the effects
of membrane tension (Sec.~\ref{s:tension}), asymmetric potential (Sec.~\ref{s:asymmetry}),
and noise (Sec.~\ref{s:noise}).

\section{Finite Membrane tension}
\label{s:tension}


Even though bending rigidity plays a major role in membrane dynamics,
experiments usually report the existence of an effective
tension, varying from $10^{-5}$ to $10^{-3}$ Jm$^{-2}$~\cite{Swain2001,Malaquin2010,Hemmerle2013}.
Therefore, in this section we wish to discuss its effect.

As already noticed in Ref.~\cite{LeGoff2014},
the tension-dominated limit where $\kappa=0$ and $\sigma\neq 0$
leads to the standard TDGL and CH models 
respectively for permeable and impermeable walls,
which exhibit monotonous kink profiles
leading to attractive non-oscillatory interaction between kinks,
which trigger perpetual coarsening.
From these results, it is natural to investigate 
the dynamics at finite values of $\sigma$ and $\kappa$,
mainly for two reasons.
First, we aim to identify the threshold above which
coarsening can be observed. Second, even
when no perpetual coarsening is present, we point out that tension
is able to affect the spatial organization of the frozen states
observed at $\sigma=0$.

\subsection{Critical tension}

As discussed above, a crucial property of Eqs.(\ref{e:TDGL4tension},\ref{e:CH4tension})
is the existence of oscillations in the kink tails, at the edges of adhesion patches.
Consider a small perturbation $H=H_{m}+\delta H$, where $H_{m}$ is a minimum of the potential $U$.
To leading order, both in the conserved and non-conserved cases, stationary states obey 
\begin{eqnarray}\label{eqdelta}
\partial_{X}^{4}\delta H-\gamma\partial_{XX}\delta H+U''_{m}\delta H=0,
\end{eqnarray}
where $U''_{m}=U''(H_{m})$. Assuming $\delta H\sim\exp(-r X)$,
we obtain
\begin{eqnarray}
r^{4}-\gamma r^{2}+U''_{m}=0.
\label{e:r}
\end{eqnarray}
Solving Eq.(\ref{e:r}) with the two constraints: $U''_m>0$, and $\gamma>0$, 
we find two different regimes separated by the critical tension
\begin{eqnarray}
\gamma_{c}=(4U''_{m})^{{1}/{2}}.
\end{eqnarray}

For small normalized tensions  $\gamma<\gamma_c$
the oscillations are still present. Considering only the profiles $\delta H(X)=R(X)$
decaying for $X\rightarrow +\infty$ (profiles decaying as $X\rightarrow -\infty$
can be obtained by symmetry), we find
\begin{equation}
R(X)=A\cos(Q_1X+\alpha)\exp(-Q_2X).
\label{e:kink_profile_oscill}
\end{equation}
where 
\begin{equation}
Q_{1,2}=\displaystyle\left[\left({U''_m}^{1/2}\right)/2\mp\gamma/4\right]^{1/2}
.
\end{equation} 
Hence, the wavelength of the oscillation $\Lambda=2\pi/Q_1$ increases when
the tension increases. At the threshold, $\Lambda$ diverges
and the kink profile becomes monotonic.

For larger normalized tensions $\gamma>\gamma_c$,
the profile is the superposition of two non-oscillating exponentials:
\begin{eqnarray}
R(X)&=&A_+\exp(-Q_+X)+A_-\exp(-Q_-X),
\label{e:Q+-}
\end{eqnarray}
where 
\begin{eqnarray}
Q_\pm=2^{-1/2}[\gamma\pm(\gamma^2-4U''_m)^{1/2}]^{1/2}.
\end{eqnarray}

Following Ref.~\cite{LeGoff2015a},
the function $\tilde R$ which intervenes in the evolution
equation~(\ref{e:kink_dynamics_TDGL_ext}-\ref{e:kink_dynamics_CH_ext}) for the kink 
positions,
is obtained from $R$ as follows:
\begin{eqnarray}
\tilde R(X)=2\left[  
 U''_m  R^2\left(\frac{X}{2}\right) 
-  R''^2\left( \frac{X}{2}\right)
\right] .
\label{e:tilde_R}
\end{eqnarray}
Hence, $\tilde R$ is oscillatory when $R$
is oscillatory.
As a consequence of the disappearance of oscillatory
kink tails in $R$ and $\tilde R$, we expect that coarsening should be restored
for $\gamma>\gamma_c$.

\begin{figure}
\includegraphics[width=1.\columnwidth]{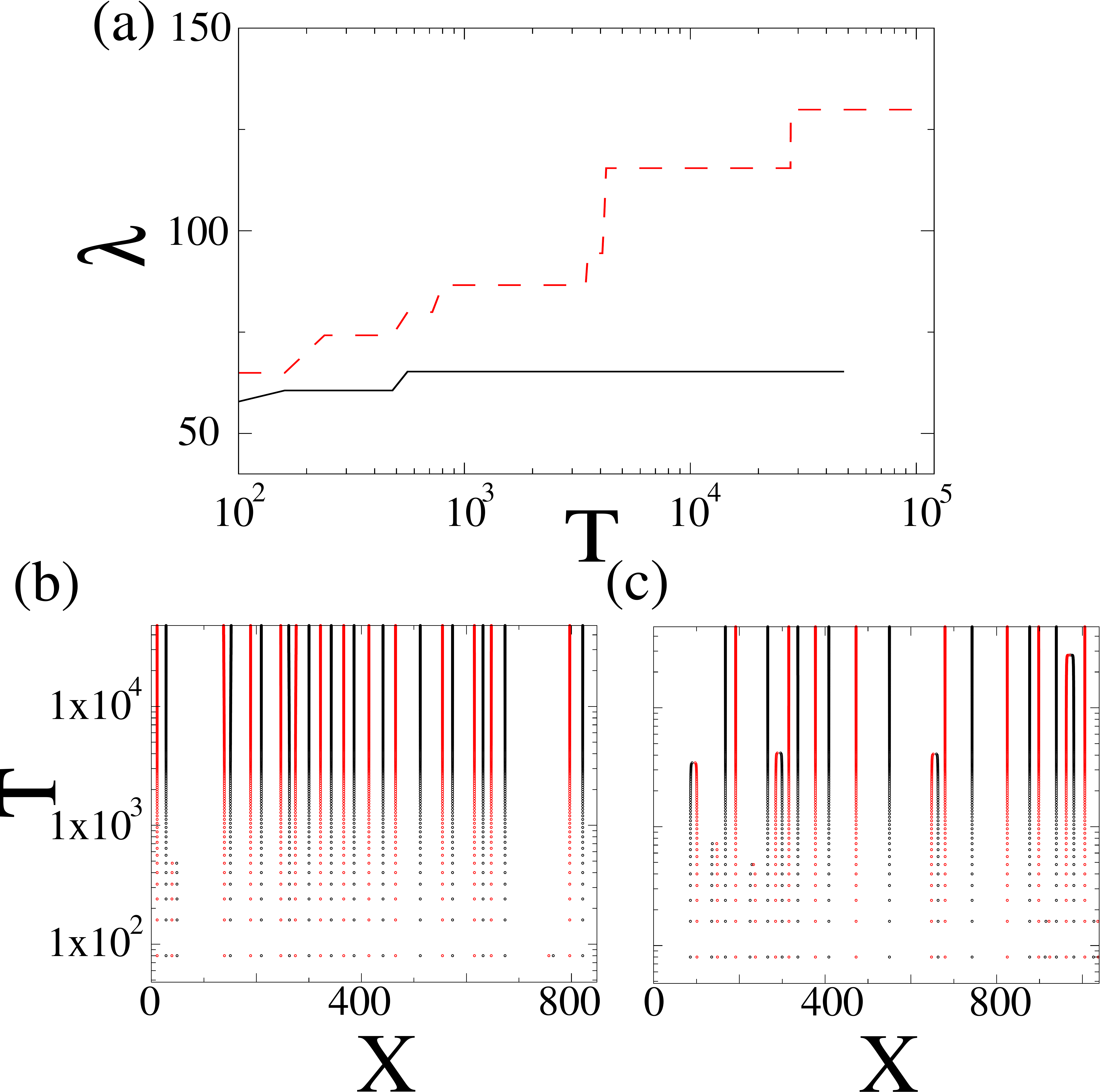}
\caption{
Non-conserved dynamics.
(a) Average wavelength as a function of time. 
Black solid line corresponds to $\gamma=2$, and the red 
dashed line to $\gamma=3$. (b) Zeros of the membrane profile for $\gamma=2$, 
and (c) for $\gamma=3$. Black points correspond to the condition $h=0$ and $\partial_xh>0$,
and red points correspond to $h=0$ and $\partial_xh<0$. 
}
\label{lambdat2}
\end{figure}

\begin{figure}
\includegraphics[width=1.\columnwidth]{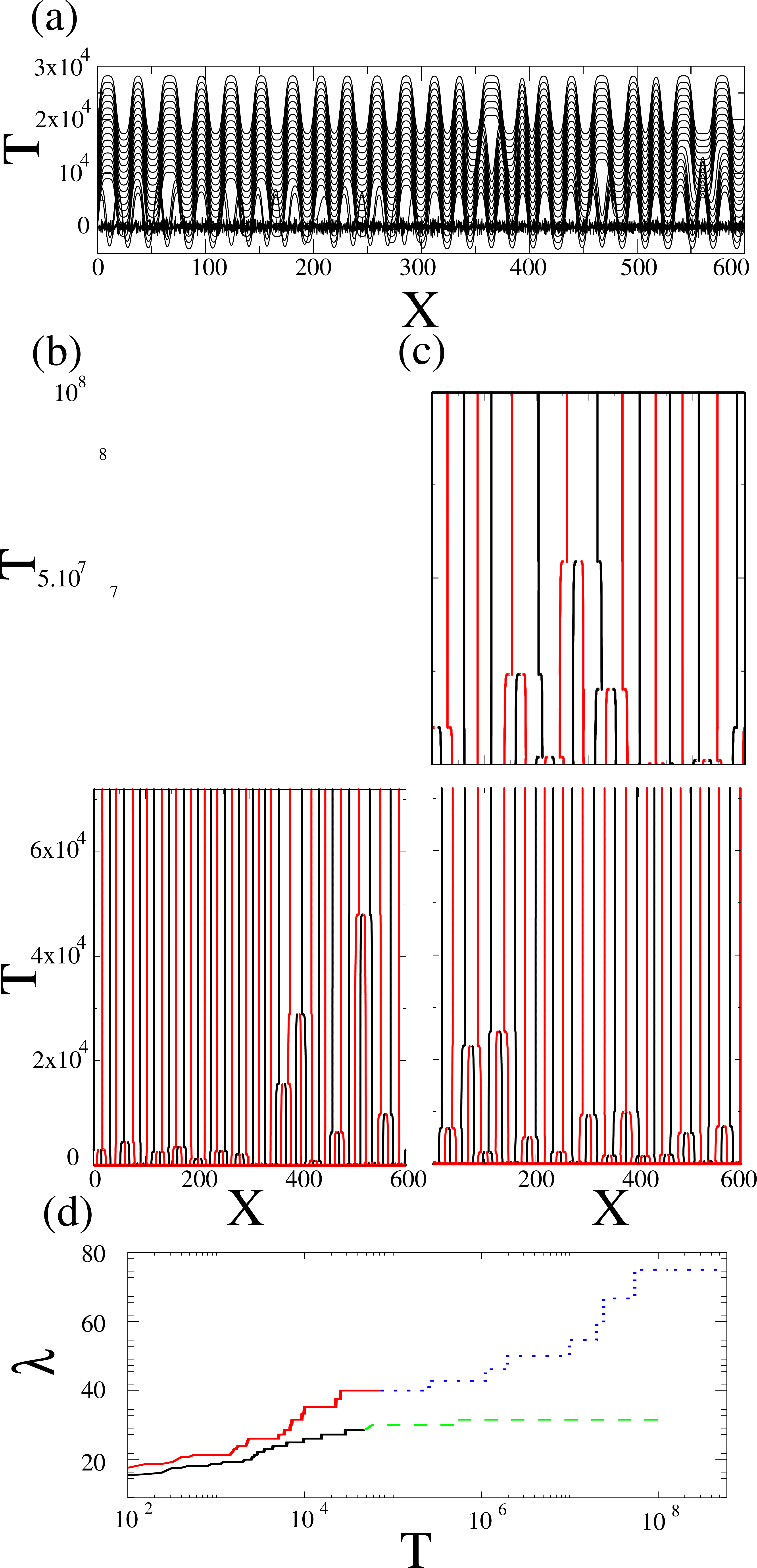}
\caption{Conserved dynamics.
(a) Membrane profile as a function of time for $\gamma=2$. 
(b) Trajectories of the zeros of the membrane profile. 
Bottom: full simulation; top: subsequent dynamics obtained from the kink model. 
(c) Same plots as (b) for $\gamma=3$.
(d) Average wavelength as a function of time.
Interrupted coarsening is found in the
lower curve with $\gamma=2$: 
the black solid line corresponds to the full dynamics, and the green dashed 
line is obtained from the kink model for $\gamma=2$. 
Endless coarsening is found in the upper curve with $\gamma=3$: 
red solid line for full dynamics and blue dotted line 
for the kink model.
}
\label{evolh2}
\end{figure}

In order to check this prediction,
we have obtained numerical solutions of Eqs.(\ref{e:TDGL4tension},\ref{e:CH4tension}) in the presence of tension
with an initial condition consisting of small perturbations around the average height $\bar H=0$.
We choose the specific potential
\begin{eqnarray} 
U(H)=\frac{1}{4}(H_m^2-H^2)^2, 
\label{e:quartic_U}
\end{eqnarray}
with $H_{m}=0.9$, leading to
the critical tension  $\gamma_{c}\simeq2.55$.
The existence of this threshold is confirmed both in  the conserved and non-conserved cases.
As shown in Figs.~\ref{lambdat2},\ref{evolh2}, coarsening is stopped for $\gamma=2$,
while perpetual coarsening  is observed for $\gamma=3$.
While the full membrane dynamics Eq.(\ref{e:TDGL4tension}) can be implemented for the non-conserved case,
we have used 
kink dynamics Eq.(\ref{e:kink_dynamics_CH_ext}) to reach long enough time-scales
in the conserved case. For $\gamma=3>\gamma_c$,
we have implemented the kink dynamics
using only the exponential contribution $Q_+$
in Eq.(\ref{e:Q+-}).
Indeed, due to its slower decay, the term involving $Q_+$ is always be dominant
at large scales.

\subsection{Transient coarsening and disordering in CH4}

In Fig.~\ref{evolh2}, we also see that, for conserved dynamics in the presence 
of a small tension $\gamma=2<\gamma_c$, a finite amount of 
coarsening can be observed. Moreover, in contrast to the 
tensionless case, some disorder
is obtained in the final configuration.

\begin{figure}
\includegraphics[width=1.\columnwidth]{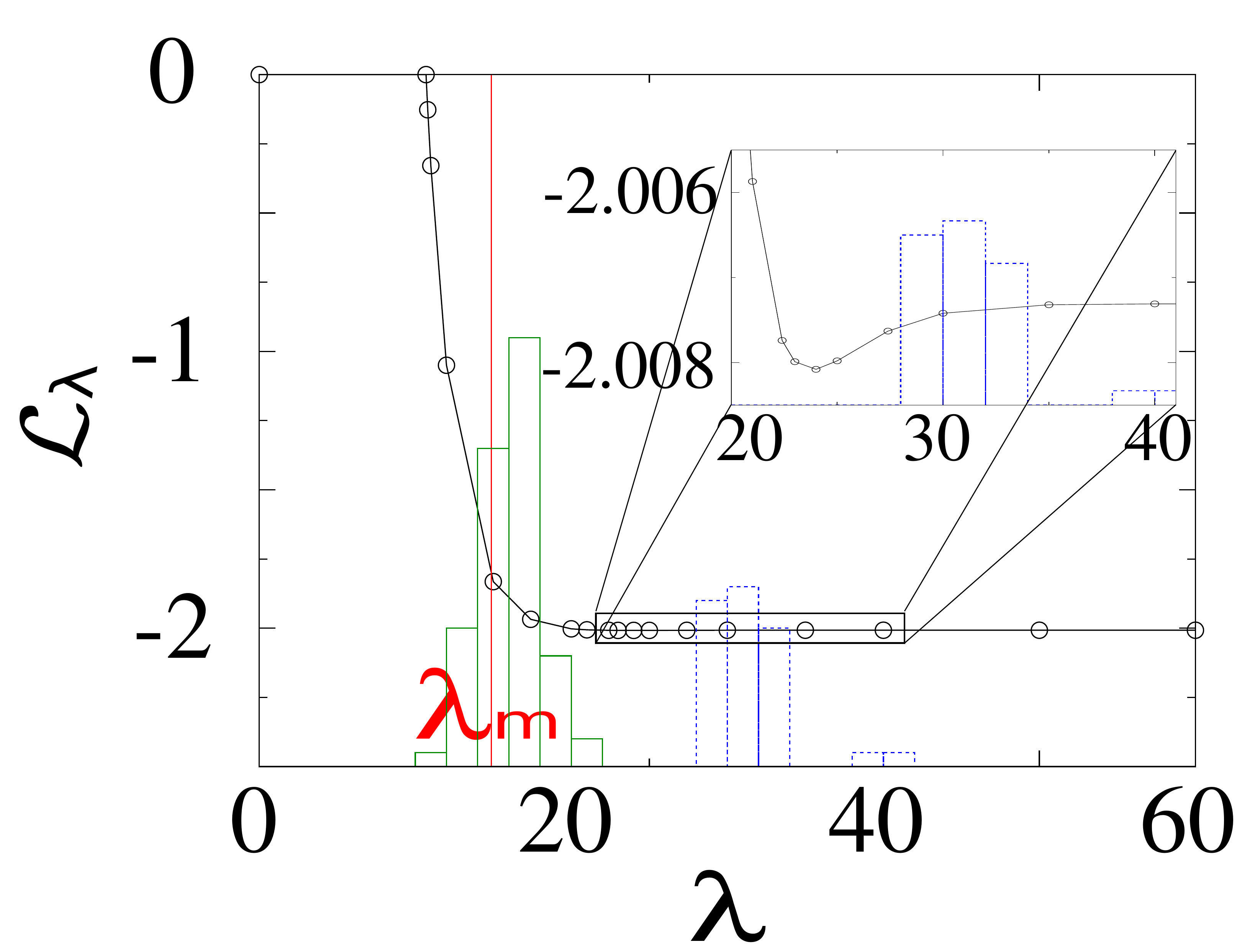}
\caption{In black with circles, $\mathcal{L}_{\lambda}$ obtained by simulations as a function of $\lambda$ 
for $\gamma=2$ and mean height $\bar{H}=0$. The red vertical line gives the position of the most 
unstable wavelength $\lambda_{m}$ in the linear analysis. The solid green histogram is the distribution 
of wavelengths for the early dynamics of the membrane with $\gamma=2$, 
and the dotted blue histogram is the long-time distribution.
\label{critere2}}
\end{figure}

To understand this result, we perform
a  straightforward extension of the results of Eq.(23) in Ref.~\cite{LeGoff2014} 
on the stability of periodic steady-states to the case of finite tensions.
We obtain that periodic steady-states are unstable if
$\partial_{\lambda}\mathcal{L}_{\lambda}\leq0$,
and stable if $\partial_{\lambda}\mathcal{L}_{\lambda}\geq0$, with
\begin{eqnarray}
\mathcal{L}_{\lambda}=-\int_{0}^{\lambda}\mathrm{d}X~\left[
2(\partial_{XX}H)^{2}+\gamma(\partial_{X}H)^{2}
\right].
\end{eqnarray}
As seen on Fig.~\ref{critere2}, this criterion reveals that the periodic state of wavelength $\lambda_m\simeq14.87$ emerging from the
linear instability is unstable, i.e.  
$\partial_{\lambda}\mathcal{L}_{\lambda_m}\leq0$. This is in contrast
with the tensionless limit where the linear instability was producing
a stable periodic steady-state~\cite{LeGoff2014}.

As a consequence of the unstable character of the periodic steady-state 
with $\lambda\simeq\lambda_m$, the system
reorganizes into a non-periodic state with a larger wavelength,
as seen in the histogram in Fig.~\ref{critere2}.
Then, the coarsening stops, and the membrane profile
is frozen. We attribute the absence of perpetual coarsening to
the existence of oscillatory interactions between the kinks,
as already discussed in Refs.~\cite{LeGoff2014,LeGoff2015a}.

\subsection{Discussion on tension effects}

In summary, tension leads to a threshold above which coarsening 
is restored, and in addition, tension can induce 
changes in the spatial organization of the adhesion patches
below the threshold. 

Using a substrate between the two gaps $\sim10$ to $20$nm,
with bending rigidity and van der Waals attraction as in Refs.~\cite{Swain2001,Hemmerle2013,LeGoff2014},
we obtain that the critical tension, 
\begin{eqnarray}
\sigma_c= \gamma_c\frac{\mathcal{U}_{0}^{1/2}\kappa^{1/2}}{h_{0}},
\end{eqnarray}
is $\sigma_c\sim 10^{-2}$Jm$^{-2}$.
This value is larger than the 
tensions $\sigma \sim 10^{-5}-10^{-3}$J.m$^{-2}$ reported by experiments~\cite{Swain2001,Malaquin2010}.
Hence, the frozen states should not be eliminated
by membrane tension in usual experimental conditions.


\section{Asymmetric potential}
\label{s:asymmetry}

Another natural extension of our model
is to consider asymmetric adhesion potentials.
Such an asymmetry occurs when
the membrane is sandwiched between two different substrates,
and therefore is expected to be the rule rather than the exception.
As an example, the adhesion of a cell membrane
sandwiched between the cytoskeleton and the extracellular matrix has no reason to be symmetric.
Moreover, model systems with vesicles and polymer brushes
exhibits a controlled asymmetric double-well potential~\cite{Bruinsma2000,Sengupta2010}.

\subsection{Asymmetric TDGL4 equation}

In order to control the asymmetry within a simple model,
we assume a potential of the form
\begin{eqnarray}
{\cal U}(h)={\cal U}_0[ U_s(H)+\beta H]
\label{e:asym_U}
\end{eqnarray}
where $U_s(-H)=U_s(H)$ is symmetric, and $\beta$ is a constant
tuning the asymmetry.
Within this model, the normalized force acting on the membrane is now:
\begin{eqnarray}
F_{Z}=-\partial_{X}^{4}H+\gamma\partial_{XX}H - U_s '(H) -\beta
\label{e:force_normalized}
\end{eqnarray}
In the following discussion on the consequences of potential asymmetry, 
we only consider the tensionless case $\gamma=0$.
As seen from Eq.(\ref{e:impermeable}) or (\ref{e:CH4tension}),
in the conserved case
this force intervenes only via its partial derivative with respect to $x$.
As a consequence, the constant term in the force, i.e. the asymmetry,
is irrelevant for impermeable walls.

\begin{figure*}
\begin{center}
\includegraphics[width=2\columnwidth]{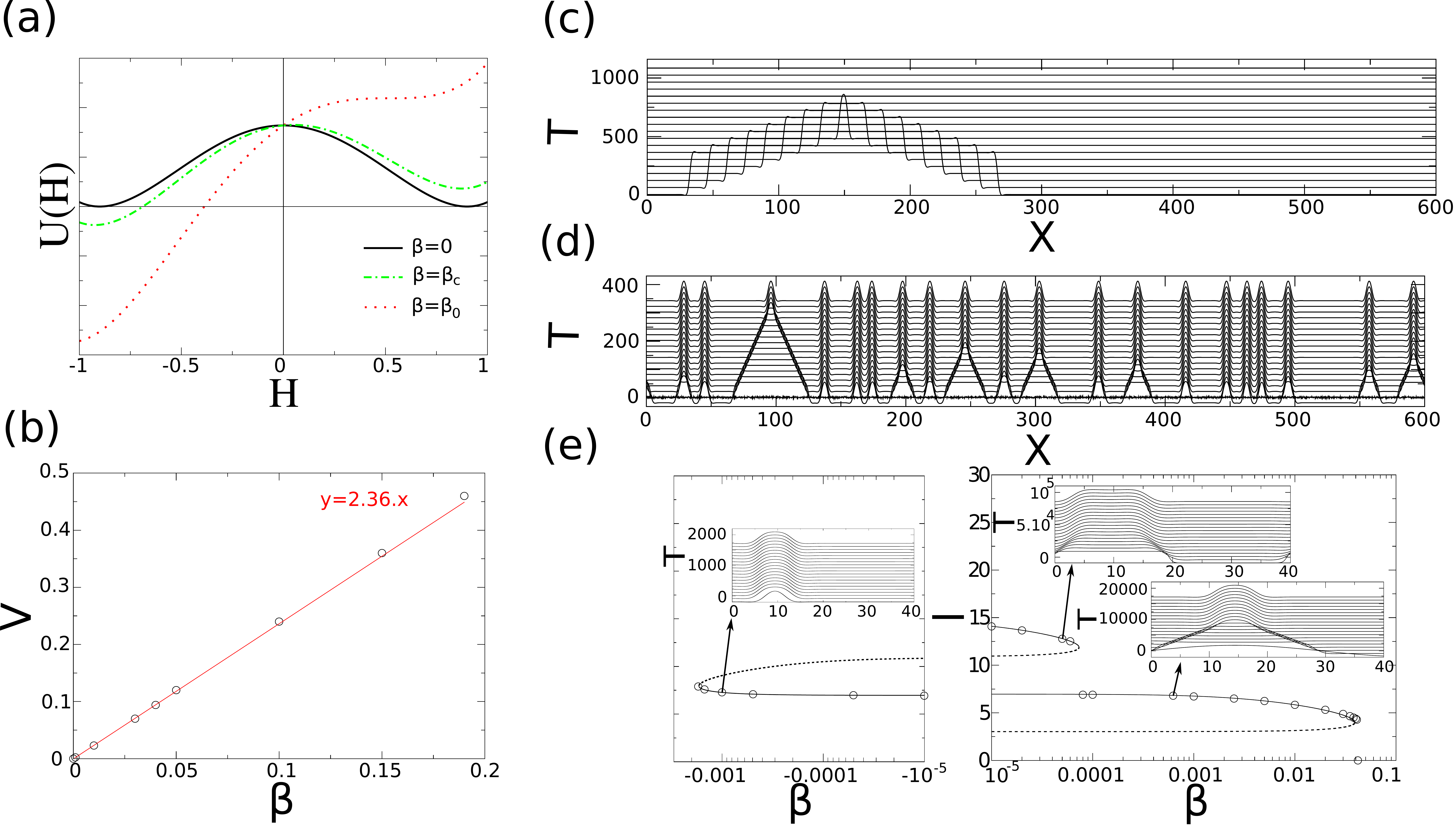}
\end{center}
\caption{
Membrane dynamics in an asymmetric double-well potential.
(a) Asymmetric potential $U(H)$ for $H_m=0.9$, 
and $\beta=\beta_0\simeq 0.281$ or $\beta=\beta_c\approx 0.041$.
(b) Velocity  $V$ of an isolated kink as a function of $\beta$. 
Circles: full simulations. The red solid line is the linear prediction
for small $\beta$, see text.
(c,d) Non-conserved dynamics in an asymmetric potential for: (c) 
$\beta=0.1$ --which is intermediate between $\beta_c$ and $\beta_0$, 
and (d) $\beta=0.04$ --which is smaller than $\beta_c$.
(e) Steady-state distance length $l$ of an isolated adhesion patch on the 
upper wall as a function of $\beta$. The circle are found 
from full simulation, and the the line is the prediction of the kink model. 
The solid line represents the stable steady-states, 
and the dashed line the unstable ones. 
The insets show the dynamics of the membrane in specific cases.
}
\label{potasy}
\end{figure*}

In contrast, the asymmetry plays an important role in 
the non-conserved case. 
In order to discuss this case in more details, we use the quartic potential Eq.(\ref{e:quartic_U}) for the symmetric part $U_s$,
and the normalized membrane evolution equation then reads:
\begin{eqnarray}
\partial_{T}H=-\partial_{X}^{4}H+H_{m}^{2}H-H^{3}-\beta.
\label{NCbeta}
\end{eqnarray}
First, note that for very strong asymmetries 
$|\beta|>\beta_0=2H_m^3/3^{3/2}\simeq 0.385 H_m^3$, there
is only one minimum, and the dynamics consists in a trivial relaxation
to a flat membrane in a single potential well.

For moderate $\beta$, i.e. $\beta_0>|\beta|>\beta_c\approx 0.056 H_m^3$,  
the numerical solution of Eq.(\ref{NCbeta}) indicates that kinks
drift at constant velocity and annihilate so that the whole membrane
moves to the lowest potential minimum in finite time,
as shown in Fig.\ref{potasy}(c).
For smaller values of the asymmetry $|\beta|<\beta_c$,
the drift of kinks is still observed, but depending on the initial conditions, the membrane
sometimes ends up in a configuration with frozen asymmetric
adhesion patches in the unfavorable well of the potential,
as shown in Fig.\ref{potasy}(d).
In the following subsections we discuss and analyze quantitatively
these results.


\subsection{Kink drift}
\label{s:kinkdrift}

Let us first discuss the kink drift.
Consider an isolated kink separating two adhesion domains
in the two potential wells. If the depth of the potential wells 
are different, the total energy can be decreased by 
a displacement of the kink in the direction that increases the size
of the adhesion domain with the lowest energy.
We assume that a kink drifts at some constant velocity $V$,
with a profile $H_k(X-Vt)$. Multiplying Eq.~\eqref{NCbeta} by $\partial_{X}H_k$ 
and integrating over $x$, we find
\begin{eqnarray}
V=\beta\frac{B_0}{B_1}.
\label{e:kink_veloc}
\end{eqnarray}
Due to the change of sign of $B_0$, kinks and antikinks drift in opposite directions
(due to symmetry, the absolute values of their drift velocities
are also equal).

In general, the kink profile $H_k$ depends
on $\beta$, so that $V$ might exhibit a complex dependence
on $\beta$ via $B_0$ and $B_1$ in the r.h.s. of Eq.~(\ref{e:kink_veloc}). 
However, we expect from Eq.~(\ref{e:kink_veloc}) that $V$ should be linear in $\beta$ when $\beta\ll 1$.
In addition, due to the $H\rightarrow-H$ symmetry of Eq.(\ref{NCbeta}) at $\beta=0$,
both $B_0=H_k(+\infty)-H_k(-\infty)$ and $B_1=\int_{-\infty}^{+\infty}\mathrm{d}X~(\partial_{X}H_k)^{2}$
depend only on $\beta^2$.
As a consequence, the first corrections to linearity in the dependence
of $V$ in $\beta$ should be cubic, and the linear approximation
could be a good approximation up to finite values of $\beta$.
This result is confirmed in Fig.~\ref{potasy}(b). Furthermore, the prefactor
of this linear relation
can be calculated from the kink profile  at $\beta=0$. Using the numerical
evaluation of the static kink profile at $\beta=0$ with $H_m=0.9$, we find 
from Eq.(\ref{e:kink_veloc}) $V\approx 2.36\beta$.
This is in good agreement with the direct numerical measurement
of the kink drift velocity at small $\beta$ in Fig.~\ref{potasy}(b).


\subsection{Asymmetric frozen adhesion patches}

To perform a systematic analysis of the asymmetric frozen patches, we 
have implemented the numerical solution of Eq.~(\ref{NCbeta}).  
Depending on the initial condition
and on $\beta$, one may obtain different final states.
Some examples are shown in Fig.~\ref{potasy}(c,d).

In order to rationalize these results, we have plotted in Fig.~\ref{potasy}(e)
the size of a single, steady domain in a large system as a function of $\beta$. 
In this figure, the steady-state branches
with $\beta>0$ represent finite-size patches in the higher
energy potential well, while the branches with
$\beta<0$ represent finite-size adhesion patches
in the lower potential well.
Each simulation point is obtained from a suitable
choice of initial condition.
The lower branch for $\beta>0$ is obtained with a sinusoidal
initial condition. The upper branch is obtained from 
an initial condition with a localized domain
formed by superposition of $\tanh$ functions.
Finally, the lower branch with $\beta>0$
was used as an initial condition to obtain
the steady-state solutions in the branch with $\beta<0$.

Globally Fig.\ref{potasy}(e) suggests the presence of
several branches of steady-state solutions
for $|\beta|<0.041$, while no steady-state patch
is observed for $|\beta|>0.041$.

These results can be described quantitatively within the kink model.
Indeed, the kink model  Eq.(\ref{e:kink_dynamics_TDGL_ext})
can be simply extended to account for asymmetry,
leading to:
\begin{eqnarray}
\dot{X}_n&=&\frac{1}{B_1}\left(\Delta \tilde{R}_{n}+\beta B_0\right).
\label{e:kink_dynamics_TDGL_ext_beta}
\end{eqnarray}
where the sign of $B_0$ alternates between $+$ for kinks
and $-$ for antikinks.
Considering a single adhesion domain on the upper wall, centered at $x=0$, 
we have a kink at $X_1=-\ell/2$,
and an antikink at $X_2=\ell/2$. We therefore find that steady-states, corresponding to $\partial_t\ell=0$, obey
\begin{eqnarray}
\beta
=-\frac{\tilde{R}(\ell)}{ |B_0|}.
\label{e:beta_delta}
\end{eqnarray}
In addition, from Eq.(\ref{e:kink_profile_oscill}):
\begin{align}
\tilde{R}(\ell)=2A^{2}U''_{m}\cos\left[\frac{\ell{U''_m}^{1/4}}{2^{1/2}}+2\alpha\right]
\exp\left[-\frac{\ell{U''_m}^{1/4}}{2^{1/2}}\right].
\label{e:tilde_R}
\end{align}
For  $\beta\ll 1$, 
the constants can be determined numerically from the symmetric case at $\beta=0$,
leading to:  $|B_0|\approx 2H_{m}$, $A=0.87$, and $\alpha=2.72$~\cite{LeGoff2014}.
As shown in Fig.~\ref{potasy}(e), 
these assumptions allow one to obtain a good quantitative agreement between the
kink model (solid and dashed lines), and the
steady-states observed in the simulations (symbols).

Furthermore, the stability of these steady-states can also be understood
within the kink model. Indeed, consider a stationary state with two kinks 
separated by the distance $\ell$. Then, assuming a perturbation of $\ell$ equal to $\delta \ell\propto\exp(i\omega T)$,
we obtain from Eq.(\ref{e:kink_dynamics_TDGL_ext_beta})
\begin{eqnarray}
i\omega=-\frac{2\tilde{R}'(\ell)}{B_1}
\end{eqnarray}
indicating that if $\tilde{R}'(\ell)\leq0$ the state is unstable.
Using Eq.(\ref{e:beta_delta}), we find that the steady-state is unstable if $\partial_{\ell}\beta\geq0$.
This is in agreement with the results of Fig.~\ref{potasy}(e),
where no steady-state is observed in the unstable branches with $\partial_{\ell}\beta\geq0$,
indicated with a dashed line.

\subsection{Discussion on asymmetry}

In summary, in the non-conserved case, a finite potential asymmetry is needed in order to 
eliminate the frozen adhesion patches. 
However, we recall that asymmetry in the depth of potential wells has no effect in the presence of impermeable
substrates.

Asymmetric two-state adhesion potentials have been obtained in
experiments~\cite{Bruinsma2000,Sengupta2010}
using competition between a medium-range repulsion created by polymer brushes (mimicking the glycocalyx
of cells), and a short range attraction resulting from the attachment of
ligand-receptor pairs.

It is difficult to provide precise quantitative predictions
for these experiments because the potentials of Refs.~\cite{Bruinsma2000,Sengupta2010}
do not exhibit the simple quartic form with a linear bias that we assumed here in Eqs.(\ref{e:quartic_U},\ref{e:asym_U}).
However, asymmetries can be compared using a simple dimensionless
parameter 
\begin{eqnarray}
b=\frac{U(H_{m+})-U(H_{m-})}{U(H_{M})-[U(H_{m+})+U(H_{m-})]/2},
\end{eqnarray}
where $H_{m\pm}$ are the
positions of the two minima of the potential, and $H_{M}$ is the
position of the maximum. 
Our model potential Eqs.(\ref{e:quartic_U},\ref{e:asym_U}) leads to 
$b\approx 8\beta/H_m^3$ for small $\beta$. As a consequence,
the critical asymmetry above which
patches cannot survive is $b_c\approx8\beta_c/H_m^3\approx 0.45$.

We find
a smaller asymmetry $b\approx 0.3$ from Fig.5 of Ref.~\cite{Bruinsma2000}, 
and a larger asymmetry $b\approx 1.5$ for Ref.~\cite{Sengupta2010},
suggesting that adhesion patches may
survive the asymmetry for the potential of Ref.~\cite{Bruinsma2000},
but not for that of Ref.~\cite{Sengupta2010}.
This conclusion should not be taken as
a quantitative statement. However, it suggests that
both situations could be observable in this type of experiments,
provided that the substrate is permeable.

 
\section{Thermal noise}
\label{s:noise}

Another important physical ingredient which is able to
affect the frozen states and
restore coarsening is thermal fluctuations.
Lipid membranes usually exhibit a bending rigidity 
of the order of 30~$k_BT$, and are therefore subject
to significant thermal fluctuations.
In this section, we investigate the consequences of thermal fluctuations
on membrane dynamics.


\subsection{Noisy kink dynamics}

 We include Langevin
forces in the kink dynamics equations following the same lines as
in Ref.~\cite{Kawasaki1982,Kawakatsu1985}. As discussed in  Appendix~\ref{a:kink_noise_amplitude_FDT}, we then have for
the non-conserved and conserved cases respectively
\begin{eqnarray}
\dot{X}_n&=&\frac{1}{B_1}\Delta \tilde{R}_{n}+\zeta_n(T),
\label{e:kink_dynamics_TDGL_ext_noisy}
\\
\dot{X}_n&=&\frac{1}{B_0^2}
\left(\frac{\tilde{R}_{n+3/2}-\tilde{R}_{n-1/2}}{\ell_{n+1/2}}
+\frac{\tilde{R}_{n+1/2}-\tilde{R}_{n-3/2}}{\ell_{n-1/2}}
\right)
\nonumber \\&&
+\frac{\xi_{n+1/2}(T)}{\ell_{n+1/2}^{1/2}}
+\frac{\xi_{n-1/2}(T)}{\ell_{n-1/2}^{1/2}},
\nonumber \\
\label{e:kink_dynamics_CH_ext_noisy}
\end{eqnarray}
where the Langevin forces $\zeta$ and $\xi$ are zero-average white Gaussian noise.
Their correlations read
\begin{eqnarray}
\langle\zeta_{n_1}(T_1)\zeta_{n_2}(T_2)\rangle= 2D_\zeta \delta_{n_1n_2}\delta(T_1-T_2)
\label{e:kink_dynamics_TDGL_ext_correl}
\\
\langle\xi_{n_1}(T_1)\xi_{n_2}(T_2)\rangle= 2D_\xi \delta_{n_1n_2}\delta(T_1-T_2)
\label{e:kink_dynamics_CH_ext_correl}
\end{eqnarray}
where $\delta_{n_1n_2}$ and $\delta(t)$ are respectively the Kronecker symbol and Dirac delta function.
In the conserved case Eq.(\ref{e:kink_dynamics_CH_ext_noisy}), 
we have neglected the subdominant terms proportional to $B_2$ 
in Eq.(\ref{e:kink_dynamics_CH_ext}), which are not expected
to affect qualitatively the asymptotic dynamics.

The noise amplitudes 
are derived in Appendix~\ref{a:kink_noise_amplitude_FDT} 
using the fluctuation-dissipation theorem:
\begin{eqnarray}
D_\zeta &=&\frac{k_BT}{B_1{\cal U}_0^{3/4}h_0^{1/2}\kappa^{1/4}}
\label{e:D_zeta}
\\
D_\xi &=& \frac{k_BT}{B_0^2{\cal U}_0^{3/4}h_0^{1/2}\kappa^{1/4}}
\label{e:D_xi}
\end{eqnarray}

We have implemented numerically these Langevin equations.
The details of the numerical scheme are described in Appendix~\ref{a:numerics}.


\subsection{Activated coarsening}

\begin{figure*}
\includegraphics[width=1.8\columnwidth]{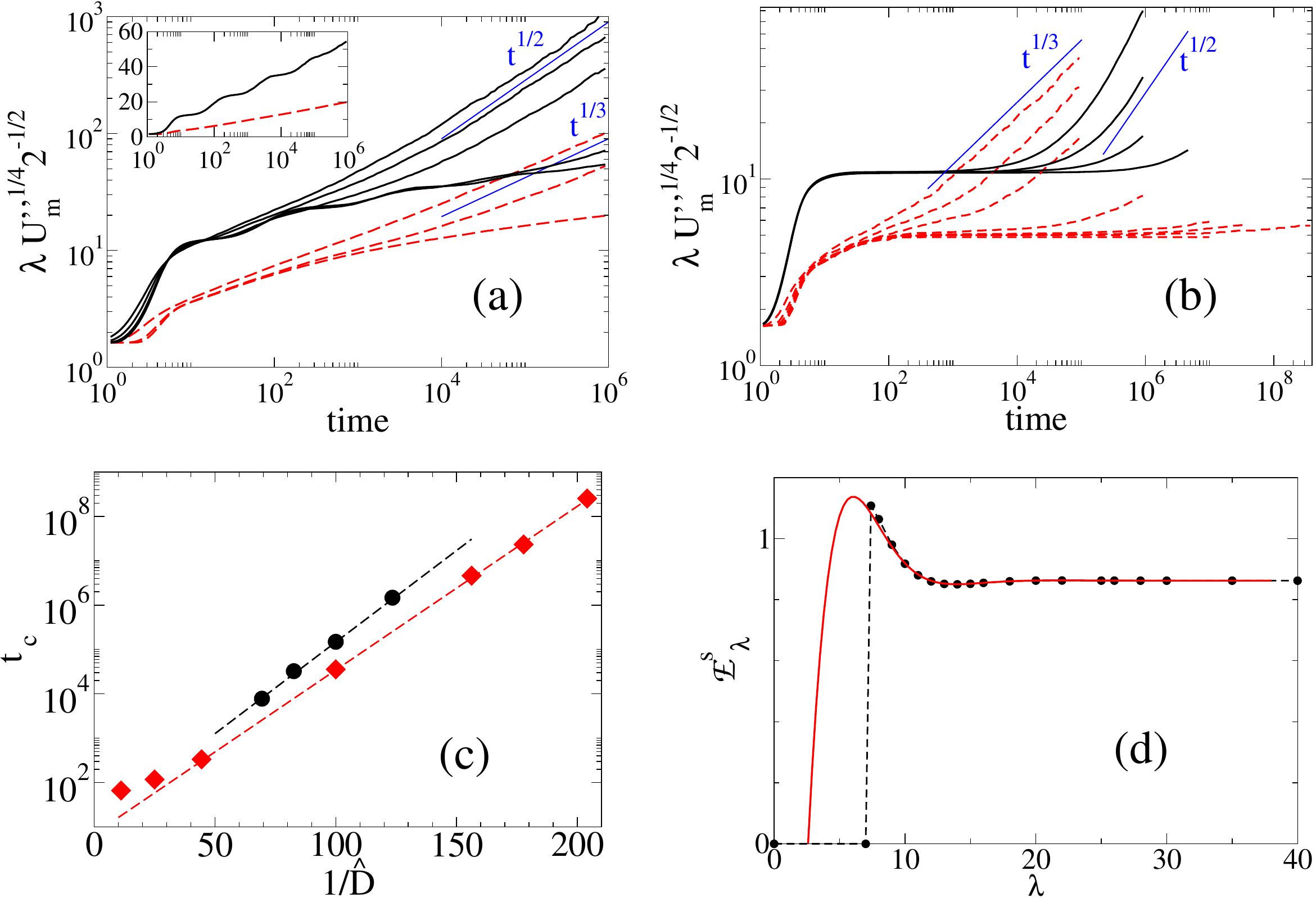}
\caption{Activated coarsening.
(a) Normalized average distance $\lambda U_m^{1/4}2^{-1/2}$ between kinks as a function of normalized time.
Noisy-TDGL kink dynamics: black solid lines, with from bottom to top  $\hat D=0.002, 0.01, 0.1, 0.2, 0.3$.
Noisy-CH kink dynamics: red dashed lines, with from bottom to top  $\hat D=0.002, 0.1, 0.3$.
Thin lines are expected power-laws at long times, 
and the inset in semi-log coordinates shows logarithmic coarsening
in the case of low noise amplitude $\hat D=0.002$ for TDGL (solid line),
and CH (dashed line).
(b) Normalized average distance $\lambda U_m^{1/4}2^{-1/2}$ between kinks
as a function of normalized time. 
Noisy-TDGL4 kink dynamics: black solid lines, with from bottom to top  $\hat D=0.09, 0.1, 0.11, 0.12$.
Noisy-CH4 kink dynamics: red dashed lines, with from bottom to top  $\hat D=0.07, 0.07 0.075, 0.08, 0.1 0.15, 0.2 0.3$.
(c) Time $t_c$ for the initiation of coarsening: discs for TDGL4,
and diamonds for CH4.
(d) Energy of a periodic steady-state ${\cal E}^s_\lambda$ as a function
of its period $\lambda$ for the potential defined in Eq.(\ref{e:quartic_U}),
with $H_m=0.9$. The symbols (dashed line) indicate the 
full numerical solution of the membrane profile. The red solid line
is the large $\lambda$ approximation
from Eq.~(\ref{e:energ_periodic_SS_large_lambda}).}
\label{fig:noise_coarsening}
\end{figure*}

The numerical solution of Eqs.(\ref{e:kink_dynamics_TDGL_ext_noisy},\ref{e:kink_dynamics_CH_ext_noisy}) indicates that thermal fluctuations
always lead to coarsening at long times.
As shown in Fig.~\ref{fig:noise_coarsening}(a,b), 
the coarsening exponent $1/2$ for noisy-TDGL4 and 
$1/3$ for noisy-CH4, are the same as those found for
noisy-TDGL and noisy-CH respectively. 
The same exponents would be observed
if the deterministic terms in r.h.s. of 
Eqs.(\ref{e:kink_dynamics_TDGL_ext_noisy},\ref{e:kink_dynamics_CH_ext_noisy}) were absent.
This suggests
that the precise form of the linear terms (second or fourth order)
is irrelevant at long times, and asymptotic coarsening is controlled
only by the noise and the conservation law.

In contrast, the short-time behavior is strongly influenced by
the deterministic stabilizing terms.
In the TDGL and CH cases, the coarsening is logarithmic in
the early time dynamics, as in the deterministic case~\cite{Kawasaki1982,Kawakatsu1985}.
Then, it crosses over to an asymptotic power-law in the late stages. 
This well known behavior is shown in Fig.~\ref{fig:noise_coarsening}(a).
Similarly, the TDGL4 and CH4 noisy kink dynamics behave like the deterministic
dynamics at short times, i.e. with arrested dynamics. 
Then, we find a crossover to the expected power-law
behavior, as seen in Fig.~\ref{fig:noise_coarsening}(b). 

The crossover time $t_c$
to the coarsening regime in TDGL4 and CH4 exhibits an exponential
dependence in the noise amplitude:
\begin{eqnarray}
t_c &=& t_{c0}\exp\left[\frac{E_0}{\hat D}\right]
\nonumber \\
&=& t_{c0}\exp\left[E_02^{1/2}A^2{U''_m}^{3/4}\frac{{\cal U}_0^{3/4}h_0^{1/2}\kappa^{1/4}}{k_BT}\right]
\label{e:tc}
\end{eqnarray}
where we have used the normalized noise amplitude 
\begin{eqnarray}
\hat D=\frac{k_BT}{2^{1/2}A^2{U''_m}^{3/4}{\cal U}_0^{3/4}h_0^{1/2}\kappa^{1/4}},
\label{e:hat_D}
\end{eqnarray}
defined in the normalized equations of Appendix~\ref{a:numerics}.

Note that, in order to obtain the dependence 
of our results on the kink parameters ($A$, $U_m''$, $B_0$,
and $B_1$), we have performed the noisy kink simulations with a special set of normalized coordinates
defined in Appendix~\ref{a:numerics}. This dependence in the kink parameters
then appears  explicitly when going back to the coordinates used in main text.

We have measured $t_c$ using
an arbitrary threshold wavelength $\lambda_c$
from the relation $\lambda(t_c)=\lambda_c$, 
where $\lambda(t)$ is the average distance between
kinks at time $t$. 
We started from randomly distributed kinks with
an initial average separation $\lambda(t=0)=2.30\times U_m''^{-1/4}$ corresponding
to the most unstable wavelength within the kink model.
We obtain $E_0\approx0.095$  for TDGL4 with $\lambda_c=17.0\times U_m''^{-1/4}$, 
and $E_0\approx0.085$ for CH4 with $\lambda_c=7.92\times U_m''^{-1/4}$.

This thermal activation of coarsening
can be intuitively understood from the need to overcome
small energy barriers corresponding to the oscillatory
interactions between kinks. 

The existence of barriers
can be quantitatively discussed within the kink approximation.
One basic assumption underlying the kink description,
and explored in details in Ref.~\cite{LeGoff2015a},
is that the profile between two kinks can be approximated by
that of a periodic steady-state.
Using this approximation, 
we may design an expression for the energy 
of a periodic steady-state with a large $\lambda$
\begin{eqnarray}
&&{\cal E}^s_\lambda
=2{\cal E}_{kink}
\nonumber\\
&+&4A^2{U''_m}^{3/4}
\sin\left[ 
\frac{\lambda{U''_m}^{1/4}}{2^{3/2}}+2\alpha-\frac{\pi}{4}
\right]
\exp\left(
\frac{-\lambda{U''_m}^{1/4}}{2^{3/2}}
\right)
\nonumber \\
\label{e:energ_periodic_SS_large_lambda}
\end{eqnarray}
where ${\cal E}_{kink}$ is the energy of an isolated kink.
The detailed derivation of this result is 
reported in Appendix~\ref{a:kink_noise_amplitude_FDT}.
A comparison to the exact energy computed numerically in Fig.~\ref{fig:noise_coarsening}(d) shows that 
Eq.(\ref{e:energ_periodic_SS_large_lambda}) is a good
approximation for the energy for large $\lambda$.
When $\lambda<8.35..\times U_m^{-1/4}$, the full numerical solution appears to be unstable.
The expression (\ref{e:energ_periodic_SS_large_lambda}) exhibits a maximum around $\lambda\approx 6.83..\times U_m^{-1/4}$,
and for smaller distances, pairs of kinks are expected to experience an attraction
leading to annihilation. The energy barrier, i.e. the difference between the minimum
energy and the maximum energy that can be reached before annihilation
is similar in both cases. Since we wish to approximate
the profile between two kinks by half a periodic steady-state,
the effective energy barrier is half the barrier observed in
Fig.~\ref{fig:noise_coarsening}(d). We find similar values $E_b^{th}\approx 0.143$
from Eq.(\ref{e:energ_periodic_SS_large_lambda}), and $E_b^{num}\approx 0.13$
from the numerical solution of the full profile.
Hence the kink model provides a reasonable description
of the energy barrier.

Using Eq.~(\ref{e:tc}) with Eqs.~(\ref{e:D_zeta},\ref{e:D_xi}),
we find the expected value of $E_0$ for both the conserved and non-conserved cases:
\begin{eqnarray}
E_0 &=& \frac{E_b}{2^{1/2} A^2 {U''_m}^{3/4}} .
\end{eqnarray}
Using the above-mentioned value $E_b^{th}\approx 0.143$, 
we finally obtain  $E_0\approx0.093$,
in good agreement with the values extracted from the exponential
dependence of $t_c$ (see above).
Hence,
Eq.(\ref{e:energ_periodic_SS_large_lambda}) provides a quantitative understanding
of the origin of the energy barriers controlling the activation
of coarsening in noisy kink dynamics.

\subsection{Discussion on noise}

The results of this section can be interpreted
qualitatively within a simple picture
where coarsening is controlled by the competition between two timescales.
The first timescale $t_{diff}\sim \lambda^\theta$ is the time needed for non-interacting
kinks moved only by Langevin forces to collide with their neighbors,
which are initially located at a typical distance $\lambda$.
From the dimensional analysis of Eqs.(\ref{e:kink_dynamics_TDGL_ext_noisy},\ref{e:kink_dynamics_CH_ext_noisy}),
one finds $\theta=2$ for the non-conserved case, and $\theta=3$ for the conserved case.

In tension-dominated models, i.e. TDGL or CH, the second time-scale is the time $t_{int}$ needed
for two neighboring kinks to annihilate due to their
deterministic mutual attraction. 
Since this attraction decreases exponentially with the distance,
we have $t_{int}\sim {\rm e}^\lambda$.
Here, kink random motion and deterministic interactions
act in parallel, and the {\it shortest} of the two timescales dominates.
Thus, initially for small $\lambda$, and if the noise strength
is small enough, one has $t_{int}\ll t_{diff}$,
leading to a dynamical behavior dominated
by $t_{int}$, with the standard logarithmic coarsening law $\lambda\sim \ln t$.
However, at large times $t_{int}\gg t_{diff}$.
As a consequence, the dynamics is dominated
by the random motion of the kinks, and one finds 
the power-law behavior $\lambda\sim t^\theta$.

In bending-dominated models, i.e. TDGL4 or CH4, the second timescale
is the time $t_c$ (given in Eq.(\ref{e:tc})) for a pair of kinks to overcome the energy barrier
for collision via thermal fluctuations.
Here, kink random motion and the passage over the energy barrier
act in series, so that the {\it largest} of the two timescales $t_{diff}$ and $t_c$ dominates.
Thus, at short times and for small enough noise strength, one has $t_{diff}\ll t_c$, and coarsening is absent.
In contrast, at long times $t_{diff}\gg t_c$, 
and the power-law coarsening with $\lambda\sim t^\theta$ is recovered.

As an important remark, the predictions 
of the previous sections on tension and asymmetry 
depend on the lengthscales $h_0$ and  $(\kappa/{\cal U}_0)^{1/2}$,
and the timescales $h_0^2\nu^{-1}{\cal U}_0^{-1}$ or $\mu\kappa^{1/2}{\cal U}_0^{-3/2}$
for the non-conserved and conserved cases respectively.
These spatial and temporal scales naturally extend to two-dimensional membranes
in three-dimensional liquids keeping the same formula and 
replacing the physical constants $\kappa$ and ${\cal U}_0$ by the two-dimensional ones
which have different dimensions. Hence,
we expect the above-mentioned results to catch
some of the physical behavior of two-dimensional membranes.

However, in the presence of thermal fluctuations,
we now have a relevant energy scale, which is 
the energy barrier ${\cal E}_b$. In physical coordinates, it reads
\begin{equation}
{\cal E}_b={\cal U}_0^{3/4}h_0^{1/2}\kappa^{1/4}E_b.
\label{e:barrier_phys}
\end{equation}
This energy scale cannot be naturally extended
to two dimensional membranes, and a simple substitution
of the energy parameters $\kappa$ and ${\cal U}_0$
by their two-dimensional counterparts provides
an expression of ${\cal E}_b$ which does not have the dimension of an energy.
Physically, confined two-dimensional membranes
would exhibit one-dimensional domain walls instead of kinks. 
The collision and annihilation of 
two one-dimensional domain walls should occur locally in a region whose spatial
extent should be fixed by the physics of the two-dimensional problem. 
It is therefore clear that we cannot directly use the result
of our model to perform quantitative predictions
about two-dimensional membranes. However,
there should still be an energy barrier for domain wall collision in two-dimensional membranes.


\section{Conclusion}
\label{s:conclusion}

In summary, we have shown that the frozen patches observed in the 1D dynamics
of membranes with bending rigidity survive up to a finite threshold
to various other physical driving forces such as tension, potential asymmetry, and thermal fluctuations.
Beyond these thresholds, coarsening is restored.
However the transition to coarsening exhibits different scenarios
in these three cases.

(i) In the presence of tension, there is a critical tension $\sigma_c$ 
above which the oscillations of the membrane profile
disappear, leading to monotonic attractive interactions
similar to that of the standard TDGL or CH equations.
The orders of magnitude indicate that the tensions usually
observed in experiments are smaller than the critical tension $\sigma_c$,
showing that frozen adhesion patches should still exist 
in typical experimental conditions.

(ii) An asymmetry in the depth of the two potential wells has no effect on the conserved case,
where walls are impermeable. However, in the non-conserved
case, i.e. for permeable walls, kinks and antikinks
experience drift forces in opposite directions, which are able
to overcome the oscillatory kink-kink interactions beyond some finite threshold.
The critical asymmetry above which frozen patches
cannot be observed is comparable to the asymmetry obtained
in experimental works~\cite{Bruinsma2000,Sengupta2010}, suggesting that 
an asymmetry-induced transition could be observed experimentally
in the presence of impermeable walls.

(iii) The presence of thermal fluctuations always lead to 
coarsening at long times. Nevertheless, the time required for the
system to undergo coarsening depends exponentially
on the noise strength, i.e. on the temperature.
Hence, for temperatures smaller that the typical energy barrier
for collision and annihilation,
the coarsening process cannot be observed.
Although we expect energy barriers to exist 
for fully two-dimensional membrane, 
we cannot conclude on the quantitative value of the barriers within
our model with a one-dimensional membrane.
 
Additional differences could appear
between one-dimensional and two-dimensional
membranes. For example, kinks in one-dimensional membranes mimic flat
domain walls separating adhesion zones between the upper and the lower
walls. However, it is clear that any effect related to the 
curvature of these domain walls cannot be accounted for within
our model with one-dimensional membranes.

To conclude, our study of idealized zero-thickness, 
and one-dimensional interfaces with bending rigidity 
sandwiched between two flat walls,
aimed at capturing qualitatively some of the complex dynamics of
lipid membranes in confined biological environments.
Our results show that bending rigidity 
is at the origin of a unique zoology of dynamical behaviors,
with finite-size patches that are robust to various physical
perturbations up to a finite threshold.


\begin{acknowledgments}
We acknowledge support from the Agence Nationale de la Recherche
Biolub Grant (ANR-12-BS04-0008).
\end{acknowledgments}


\begin{appendix}

\section{Derivation of the noise amplitude in the kink model for thermal noise}
\label{a:kink_noise_amplitude_FDT}

Here, we relate kink dynamics to an energetic picture.
The amplitude of thermal noise then follows
directly from this derivation using the 
fluctuation-dissipation theorem.

\subsection{Energy and force acting on a kink}

We start by  
decomposing the membrane into regions.
The region $n+1/2$ is located  betweens kinks $n$ and $n+1$.
The total energy
\begin{eqnarray}
{\cal E}=\int dx \left[ 
\frac{\sigma}{2}(\partial_{x}h)^2
+\frac{\kappa}{2}(\partial_{xx}h)^2
+U(h)\right],
\end{eqnarray}
is then equal to the sum of the corresponding energy contributions ${\cal E}_{n+1/2}$:
\begin{eqnarray}
{\cal E}=\sum_{n}{\cal E}_{n+1/2}.
\end{eqnarray}

In the kink picture, the energy  ${\cal E}_{n+1/2}$
is approximated by half the energy ${\cal E}^s_{\lambda}$ of a periodic steady-state
of wavelength $\lambda=2l_{n+1/2}$, where $l_{n+1/2}=x_{n+1}-x_n$.
This reads
\begin{eqnarray}
{\cal E}_{n+1/2}=\frac{1}{2}{\cal E}^s_{\lambda}|_{\lambda=2l_{n+1/2}}.
\end{eqnarray}

The energy ${\cal E}_{n+1/2}$ then only depends on the positions
of the neighboring kinks at $x_n$ and $x_{n+1}$, the force experienced
upon the motion of the $n$-th kink is
\begin{eqnarray}
F_n &=&-\frac{d}{d x_n}{\cal E}
\nonumber \\ &=&
\partial_{l_{n+1/2}}{\cal E}_{n+1/2}
-\partial_{l_{n-1/2}}{\cal E}_{n-1/2}
\nonumber \\ &=&
\partial_{\lambda}{\cal E}^s_{\lambda}|_{\lambda=2l_{n+1/2}}
-\partial_{\lambda}{\cal E}^s_{\lambda}|_{\lambda=2l_{n-1/2}}.
\end{eqnarray}
In addition, from the expression
\begin{eqnarray}
{\cal E}^s_{\lambda}=
\int_0^\lambda dx \left[ 
\frac{\sigma}{2}(\partial_{x}h)^2
+\frac{\kappa}{2}(\partial_{xx}h)^2
+U(h)\right],
\end{eqnarray}
we find
\begin{eqnarray}
\partial_{\lambda}{\cal E}^s_{\lambda}=
\frac{1}{\lambda}\int_0^\lambda dx \left[ 
-\frac{\sigma}{2}(\partial_{x}h)^2
-\frac{3\kappa}{2}(\partial_{xx}h)^2
+U(h)\right].
\nonumber \\
\label{e:deriv_energ_periodic_SS}
\end{eqnarray}

Moreover,
one can easily check that
\begin{eqnarray}
\partial_{x}h\frac{\delta E}{\delta h}
=\partial_x\left[ 
\frac{\sigma}{2}(\partial_{x}h)^2
-\kappa\partial_xh\partial_{xxx}h
+\frac{\kappa}{2}(\partial_{xx}h)^2
-U(h)\right].
\nonumber \\	
\end{eqnarray}

A periodic steady-state $h^s(x)$
by definition obeys ${\delta E}/{\delta h}=0$, and
\begin{eqnarray}
-\frac{\sigma}{2}(\partial_{x}h^s)^2
+\kappa\partial_xh\partial_{xxx}h^s
-\frac{\kappa}{2}(\partial_{xx}h^s)^2
+U(h^s)=U^*,
\nonumber \\
\label{e:def_U^*}
\end{eqnarray}
where $U^*$ is a constant.
As a consequence, Eq.(\ref{e:deriv_energ_periodic_SS})
can be rewritten as
\begin{eqnarray}
\partial_{\lambda}{\cal E}^s_{\lambda}= U^*.
\label{e:Es_U*}
\end{eqnarray}

The relation (\ref{e:Es_U*}) can be used to express the force of the $n$th kink as
\begin{eqnarray}
F_n = U^*_{\lambda}|_{\lambda=2l_{n+1/2}}
-U^*_{\lambda}|_{\lambda=2l_{n-1/2}}.
\end{eqnarray}
Since Eq.(\ref{e:def_U^*}) is valid everywhere in a steady-state,
we can evaluate it in the zone far away from kinks, where
\begin{eqnarray}
h\approx h_0
&&\left\{
H_M+R[(x-x_n)({\cal U}_0/\kappa h_0^2)^{1/4}]
\right.
\nonumber \\
&&\left.
+R[(x_{n+1}-x)({\cal U}_0/\kappa h_0^2)^{1/4}]
\right\},
\end{eqnarray}
leading to
\begin{eqnarray}
U^*_{\lambda}|_{\lambda=2l_{n+1/2}}=
{\cal U}_0\,\tilde R[l_{n+1/2}({\cal U}_0/\kappa h_0^2)^{1/4}].
\label{e:U*_and_tildeR}
\end{eqnarray}
where $\tilde R$ is defined in Eq.(\ref{e:tilde_R}).

Combining Eq.(\ref{e:U*_and_tildeR}) with Eq.(\ref{e:Es_U*}), 
one can compute a convenient expression of the energy
of a periodic steady-state:
\begin{eqnarray}
{\cal E}_\lambda=2{\cal E}_{kink}
-{\cal U}_0\int_\lambda^\infty \!\!\!dx \; 
\tilde R\left[\frac{\lambda}{2}\,\frac{{\cal U}_0^{1/4}}{\kappa^{1/4} h_0^{1/2}}\right].
\label{e:Es_tildeR}
\end{eqnarray}
which leads to Eq.(\ref{e:energ_periodic_SS_large_lambda}).

\subsection{Non-conserved dynamics}

Assuming a simple constant kink mobility $\eta$ (local in space and with no memory effect),
we write that the kink velocity is proportional to the force plus a noise term
\begin{eqnarray}
\dot x_n=\eta F_n + \bar\zeta_n= \eta {\cal U}_0 \Delta \tilde R_{n}+\bar\zeta_n.
\label{e:Langevin_kink_dyn_phys}
\end{eqnarray}
Here, $\bar\zeta_n$ is a white noise obeying
\begin{eqnarray}
\langle\bar\zeta_{n_1}(t_1)\bar\zeta_{n_2}(t_2)\rangle= 2{\cal D}_{\bar\zeta} \delta_{n_1n_2}\delta(t_1-t_2).
\label{e:kink_dynamics_TDGL_ext_correl_phys}
\end{eqnarray}
where ${\cal D}_{\bar\zeta}$ is a constant.

Comparing the deterministic part of Eq.(\ref{e:Langevin_kink_dyn_phys})
with Eq.(\ref{e:kink_dynamics_TDGL_ext_noisy}), one finds
\begin{eqnarray}
\eta=\frac{\nu \kappa^{1/4}}{2B_1 {\cal U}_0^{1/4} h_0^{3/2}}.
\end{eqnarray}
We may then use the fluctuation-dissipation
theorem, here in the form of an Einstein relation, leading to
\begin{eqnarray}
{\cal D}_{\bar\zeta}=\eta k_BT=\frac{\nu \kappa^{1/4}}{2B_1 {\cal U}_0^{1/4} h_0^{3/2}}k_BT.
\end{eqnarray} 

Finally, the normalized noise $\zeta$ used in Eq.(\ref{e:kink_dynamics_TDGL_ext_noisy})
of the main text is
related to the noise $\bar\zeta$ in physical variables via the relation
\begin{eqnarray}
\bar\zeta_n=\frac{\nu \kappa^{1/4} {\cal U}_0^{3/4}}{2 h_0^{3/2}}\zeta_n.
\end{eqnarray}

\subsection{Conserved dynamics}

In the conserved case, one starts with the observation
that, due to mass conservation, the elementary event
is the translation of a whole domain instead of that of a single kink.
The total force relative to the translation of the domain $n+1/2$
is the sum $F_{n+1}+F_n$ of the forces acting on the two kinks $n$ and $n+1$.
This translational motion is physically realized by a flux $j_{n+1/2}$
of liquid under the membrane,
to which one associates a mobility $\mu_{n+1/2}$ and a noise $\psi_{n+1}$.
The flux $j_{n+1/2}$ produces a contribution to the motion of the domain $n+1/2$
with the velocity $j_{n+1/2}/|B_0|$. Hence
\begin{eqnarray}
\frac{j_{n+1/2}}{|B_0|}=\mu_{n+1/2}(F_{n+1}+F_n)+\psi_{n+1/2},
\end{eqnarray}
with the noise correlation
\begin{eqnarray}
\langle\psi_{n_1}(t_1)\psi_{n_2}(t_2)\rangle= 2{\cal D}_{\psi,n_1}\delta_{n_1n_2}\delta(t_1-t_2),
\label{e:kink_dynamics_psi_correl}
\end{eqnarray}
where $n_1$ and $n_2$ are half-integers, and ${\cal D}_{\psi,n_1}$ depends on $n_1$
but not on $t$.

Mass conservation then allows one to
obtain the kink velocity from the fluxes
\begin{eqnarray}
\dot x_n=\frac{j_{n+1/2}+j_{n-1/2}}{|B_0|},
\end{eqnarray}
leading to Eq.(\ref{e:kink_dynamics_CH_ext_noisy}),
with the identification
\begin{eqnarray}
\mu_{n+1/2}=\frac{h_0^{1/2}{\cal U}_0^{1/4}}{24\mu B_0^2\kappa^{1/4}\ell_{n+1/2}}=\frac{h_0}{24\mu B_0^2 l_{n+1/2}},
\end{eqnarray}
where we use $l_{n+1/2}$ for physical lengths and $\ell_{n+1/2}$ for normalized lengths.

Hence, from the fluctuation-dissipation theorem
\begin{eqnarray}
{\cal D}_{\psi,n+1/2}=\mu_{n+1/2}k_BT=\frac{h_0}{24\mu B_0^2 l_{n+1/2}}k_BT.
\end{eqnarray}
Then, defining $\bar\xi_{n+1/2}=l_{n+1/2}^{1/2}\psi_{n+1/2}$ with the correlations
\begin{eqnarray}
\langle\bar\xi_{n_1}(t_1)\bar\xi_{n_2}(t_2)\rangle= 2{\cal D}_{\bar\xi}\delta_{n_1n_2}\delta(t_1-t_2),
\label{e:kink_dynamics_psi_correl}
\end{eqnarray}
we obtain a constant noise amplitude:
\begin{eqnarray}
{\cal D}_{\bar\xi}=l_{n+1/2}{\cal D}_{\psi,n+1/2}=\frac{h_0}{24\mu B_0^2}k_BT.
\end{eqnarray}

The normalized noise $\xi$ used in Eq.(\ref{e:kink_dynamics_CH_ext_noisy})
of the main text is
related to the noise $\bar\xi$ in physical variables via 
\begin{eqnarray}
\bar\xi_{n+1/2}=\frac{h_0^{3/4}{\cal U}_0^{9/8}}{24\mu\kappa^{1/8}}\xi_{n+1/2}.
\end{eqnarray}

\section{Numerical schemes for the implementation of noisy kink dynamics}
\label{a:numerics}

For the kink simulations with noise, we have further normalized all variables
in order to have all numerical prefactors in the kink equations 
equal to one. For any variable $A$, we associate a normalized
simulation variable $\hat A$.
We have therefore defined the spatial coordinate $\hat X=({U''_m}^{1/4}/2^{1/2})X$, 
the time coordinate $\hat{T}=(2^{1/2}A^2{U''_m}^{5/4}/B_1)T$,
and the noise amplitude 
$\hat{D}=B_1D_\zeta/(2^{1/2}A{U''_m}^{3/4})$
for the non-conserved case. 
In the conserved case we use the same spatial coordinate,
but different normalizations for time and noise amplitude:
the time coordinate is $\hat{T}=A^2{U''_m}^{3/2}/(B_0^2)T$,
and the noise amplitude is 
$\hat{D}=B_0^2D_\xi/(2^{1/2}A{U''_m}^{3/4})$.
Using these coordinates, the kink model equation read
\begin{eqnarray}
\dot{\hat X}_n&=&\Delta {\hat R}_{n}+\hat\zeta_n(T),
\label{e:kink_dynamics_TDGL_ext_noisy_num}
\\
\dot{\hat X}_n&=&
\left(\frac{\hat{R}_{n+3/2}-\hat{R}_{n-1/2}}{\hat\ell_{n+1/2}}
+\frac{\hat{R}_{n+1/2}-\hat{R}_{n-3/2}}{\hat\ell_{n-1/2}}
\right)
\nonumber \\&&
+\frac{\hat \xi_{n+1/2}(T)}{\hat \ell_{n+1/2}^{1/2}}
+\frac{\hat \xi_{n-1/2}(T)}{\hat \ell_{n-1/2}^{1/2}},
\nonumber \\
\label{e:kink_dynamics_CH_ext_noisy_num}
\end{eqnarray}
where 
\begin{eqnarray}
\hat{R}(\hat\ell)=\cos(\hat\ell+2\alpha) \exp(-\hat\ell).
\label{e:hat_R}
\end{eqnarray}
and the Langevin forces $\hat\zeta$ and $\hat\xi$ are zero-average white Gaussian noise.
Their correlations read
\begin{eqnarray}
\langle\hat \zeta_{n_1}(\hat T_1)\hat \zeta_{n_2}(\hat T_2)\rangle= \hat D\delta_{n_1n_2}\delta(\hat T_1-\hat T_2),
\label{e:kink_dynamics_TDGL_ext_correl_num}
\\
\langle\hat\xi_{n_1}(\hat T_1)\hat\xi_{n_2}(\hat T_2)\rangle= \hat D \delta_{n_1n_2}\delta(\hat T_1-\hat T_2).
\label{e:kink_dynamics_CH_ext_correl_num}
\end{eqnarray}

Equations~(\ref{e:kink_dynamics_TDGL_ext_noisy_num}) and (\ref{e:kink_dynamics_CH_ext_noisy_num})
are re-written as evolution equations for the 
interkink distances $\hat \ell_{n+1/2}=\hat X_{n+1}-\hat X_n$. 
The resulting equations have the form
\begin{eqnarray}
\dot{\hat\ell}_{n+1/2} &=& u_n + {\hat\zeta}_{n+1} -{\hat\zeta}_{n} 
\nonumber \\
 \dot{\hat \ell}_{n+1/2} &=& v_n+
\frac{\hat \xi_{n+3/2}}{\hat \ell_{n+3/2}^{1/2}}
+\frac{\hat \xi_{n+1/2}}{\hat \ell_{n+1/2}^{1/2}}
-\frac{\hat \xi_{n+1/2}}{\hat \ell_{n+1/2}^{1/2}}
-\frac{\hat \xi_{n-1/2}}{\hat \ell_{n-1/2}^{1/2}} ,\nonumber
\end{eqnarray}
where $u_n,v_n$ are deterministic terms. These equations
have been discretized with a standard Euler scheme, as follows
\begin{eqnarray}
&&{\hat\ell}_{n+1/2}(T+dT)= {\hat\ell}_{n+1/2}+ dT\, u_n + \sqrt{dT}\left( 
{\tilde\zeta}_{n+1} -{\tilde\zeta}_{n}\right) \nonumber \\
&& {\hat \ell}_{n+1/2}(T+dT)={\hat\ell}_{n+1/2}+ dT\, v_n+\nonumber\\ 
&& \qquad \sqrt{dT}\left(
\frac{\tilde \xi_{n+3/2}}{\hat \ell_{n+3/2}^{1/2}}
+\frac{\tilde \xi_{n+1/2}}{\hat \ell_{n+1/2}^{1/2}}
-\frac{\tilde \xi_{n+1/2}}{\hat \ell_{n+1/2}^{1/2}}
-\frac{\tilde \xi_{n-1/2}}{\hat \ell_{n-1/2}^{1/2}}\right) ,\nonumber
\end{eqnarray}
where all the quantities on the r.h.s. are calculated
at time $T$, and  $\tilde\zeta_n,\tilde\xi_n$ are gaussian random variables.

The integration time step $dT$ has been chosen as the minimum between $d\tau$ and $dT^*$,
where $d\tau$ is a fixed time step, and $dT^*=\min_n (dT^*_{n+1/2})$, where $dT^*_{n+1/2}$
is the extrapolated closure time of interval $(n+1/2)$.
This criterion allows one to have no kink annihilation, or one single
annihilation event per update. The former case occurs if $dT=d\tau$, while the latter
occurs if $dT=dT^*_{n^*+1/2}$ (in which case kinks $n^*$ and $n^*+1$ annihilate).

\end{appendix}

%

\end{document}